\documentclass[a4paper,11pt]{article}
\pdfoutput=1 

\usepackage{subcaption}
\usepackage{jcappub} 

\usepackage[T1]{fontenc} 
\usepackage{dsfont}
\def\be{\begin{equation}}
\def\ee{\end{equation}}
\def\bea{\begin{eqnarray}}
\def\eea{\end{eqnarray}}
\def\beq{\begin{eqnarray}}
\def\eeq{\end{eqnarray}}
\def\bas{\begin{subequations}\begin{eqnarray}}
\def\eas{\end{eqnarray}\end{subequations}}

\def\lsim{\mathrel{\mathop  {\hbox{\lower0.5ex\hbox{$\sim$}
\kern-0.8em\lower-0.7ex\hbox{$<$}}}}}
\def\gsim{\mathrel{\mathop  {\hbox{\lower0.5ex\hbox{$\sim$}
\kern-0.8em\lower-0.7ex\hbox{$>$}}}}}

\def\be{\begin{equation}}
\def\ee{\end{equation}}
\def\bea{\begin{eqnarray}}
\def\eea{\end{eqnarray}}
\def\beq{\begin{eqnarray}}
\def\eeq{\end{eqnarray}}
\def\bas{\begin{subequations}\begin{eqnarray}}
\def\eas{\end{eqnarray}\end{subequations}}

\newcommand{\cE}{{\mathcal E}}

\newcommand{\cH}{{\mathcal H}}

\newcommand{\cS}{{\mathcal S}}

\newcommand{\bes}{\begin{eqnarray}}
\newcommand{\ees}{\end{eqnarray}}

\def\dd{\mathrm{d}}

\numberwithin{equation}{section}

\title{\boldmath Bouncing compact objects I:  Quantum extension of the Oppenheimer-Snyder collapse}

\author[a,1]{Jibril Ben Achour,\note{Corresponding author.}}
\author[b]{Suddhasattwa Brahma,}
\author[c]{\\Jean-Philippe Uzan}

\affiliation[a]{Yukawa Institute for Theoretical Physics, Kyoto University, 606-85502, Kyoto, Japan}
\affiliation[b]{Department of Physics, McGill University, Montr\'eal, QC H3A 2T8, Canada}
\affiliation[c]{Institut d'Astrophysique de Paris, CNRS UMR 7095, Universit\'e Pierre et Marie Curie - Paris VI, 98 bis Boulevard Arago, 75014 Paris, France \\
           Sorbonne Universit\'es, Institut Lagrange de Paris, 98 bis, Boulevard Arago, 75014 Paris, France}

\emailAdd{jibril.benachour@yukawa.kyoto-u.ac.jp, suddhasattwa.brahma@gmail.com, uzan@iap.fr}

\abstract{This article proposes a generalization of the Oppenheimer-Snyder model which describes a bouncing compact object. The corrections responsible for the bounce are parameterized in a general way so as to remain agnostic about the specific mechanism of singularity resolution at play. It thus develops an effective theory based on a thin shell approach, inferring generic properties of such a UV complete gravitational collapse. The main result comes in the form of a strong constraint applicable to general UV models :  if the dynamics of the collapsing star exhibits a bounce, it always occurs below, or at most at the energy threshold of horizon formation, so that only an instantaneous trapping horizon may be formed while a trapped region never forms.  This conclusion relies solely on i) the assumption of continuity of the induced metric across the time-like surface of the star and ii) the assumption of a classical Schwarzschild geometry describing the (vacuum) exterior of the star. In particular, it is completely independent of the choice of corrections inside the star which leads to singularity-resolution. The present model provides thus a general framework to discuss bouncing compact objects, for which the interior geometry is modeled either by a classical or a quantum bounce. In the latter case, our no-go result regarding the formation of trapped region suggests that additional structure, such as the formation of an inner horizon, is needed to build consistent models of matter collapse describing black-to-white hole bounces. Indeed, such additional structure is needed to keep quantum gravity effects confined to the high curvature regime, in the deep interior region, providing thus a new challenge for current constructions of quantum black-to-white hole bounce models.\note{YITP-20-52}}
\begin{document}
\maketitle

\newpage

\section{Introduction} 
Gravitational collapse is probably the physical mechanism where General Relativity (GR) manifests the most striking deviations compared to Newtonian gravity through the formation of horizons and singularities. However, a full understanding of the final stage of the collapse requires a quantum theory of gravity in order to capture the whole evolution of the initial data. How much the UV-complete picture of gravitational collapse descending from a quantum theory of gravity will deviate from the classical description is still an open question. A crucial question is at which energy scale do quantum gravity effects become relevant. Are they confined to the deep interior, as usually assumed, or can they ``leak out'' to affect the geometry even at large distances from the center, say, for instance at the horizon scale? While the former point of view appears as the most conservative one, the second possibility has received growing interest from various approaches to quantum gravity, see for examples \cite{Giddings:1992hh, Giddings:2016btb, Giddings:2019jwy, Almheiri:2012rt, Almheiri:2013hfa, Mazur:2004fk, Mathur:2008nj, Maldacena:2013xja, Barrau:2019swg}. However, the lack of a well-defined consistent quantum theory of gravity prohibits one from exploring such questions directly. Instead, the common strategy has been to develop effective models to discuss some of these aspects with the hope that the conclusions would reveal to be robust enough when embedded in the full theory; see e.g. the reviews~\cite{Carballo-Rubio:2018jzw, Malafarina:2017csn, Barrau:2018rts}. Following this strategy, this work presents a new effective approach, independent of the details of specific quantum gravity proposals, which allows for the description of a bouncing compact object and addresses some of the above questions.

The first model of gravitational collapse was proposed almost eighty years ago by Oppenheimer and Snyder (OS) \cite{Oppenheimer:1939ue}. It models an idealized star by gluing a spatially closed Friedmann-Lema\^{\i}tre (FL) universe, filled up with a homogeneous dust, with the (vacuum) Schwarzschild exterior geometry. The dynamical cosmological interior and the Schwarzschild static exterior are glued using the Israel-Darmois junction conditions of GR. This model admits two branches, one describing a collapsing star disappearing behind a trapping horizon and eventually forming a trapped region and a singularity, i.e. a black hole. The second branch, much less acknowledged, describes an expanding ball of dust emerging from a white hole horizon, to finally reach its maximal radius.

The idea that, at the quantum level, the black hole and white hole sectors of the OS model, classically disconnected, could eventually fuse to provide a black-to-white hole quantum transition or simply a bouncing star without horizon formation, has attracted a lot of attention, motivated by results in quantum cosmology \cite{Bojowald:2015iga}. Indeed, the quantization of cosmological backgrounds has provided explicit results in this direction, realizing concrete models where the expanding and contracting branches mix to finally describe a quantum bouncing universe. Related findings on the quantization of spherically symmetric background and self-gravitating null shells confirm such results, leading to a singularity free dynamics, see Refs.~\cite{Frolov:1979tu, Frolov:1981mz, Hajicek:1992mx, Hajicek:2001yd} for early results on this aspect and Refs. \cite{Schmitz:2019jct, Kwidzinski:2020xyd} for more recent investigations. 

Among these different effective models of bouncing compact objects, the heuristic ``Planck star'' idea, introduced by Rovelli and Vidotto in Ref.~\cite{Rovelli:2014cta}, has triggered interesting phenomenological investigations \cite{DeLorenzo:2014pta, Barrau:2014hda, Barrau:2015uca, Barrau:2016fcg, Rovelli:2017zoa, Barrau:2018kyv, Rama:2019hdg}. A more detailed effective model, dubbed black hole firework, was then constructed by Haggard and Rovelli in Ref.~\cite{Haggard:2014rza} while an alternative construction was proposed around the same time by Barcelo {\em et al.} in Refs.~\cite{Barcelo:2014npa, Barcelo:2014cla}. More detailed investigations of these models were discussed in Refs.~\cite{DeLorenzo:2015gtx, DAmbrosio:2018wgv, Rovelli:2018cbg} as well as in Refs.~\cite{Barcelo:2015uff, Barcelo:2015noa, Barcelo:2016hgb, Barcelo:2017lnx}. A major prediction of these two classes of models is\footnote{In order to be trusted, this transition amplitude for the black-to-white hole tunneling has to be derived in a fully non-perturbative quantum framework. Efforts in this direction have been presented in Refs.~\cite{Christodoulou:2016vny, Christodoulou:2018ryl} where the computation of the transition amplitude is performed within the context of the spinfoam formalism of Loop Quantum Gravity (LQG). Yet, more work is needed to obtain a robust result, as important assumptions are used to set up this concrete non-perturbative calculation. More recently, an alternative approach, based on the framework of Quantum Reduced Loop Gravity, was discussed in Refs.~\cite{Alesci:2018loi, Alesci:2019pbs}.} a timescale for the bounce which is shorter than the Hawking evaporation time. As a result, the bouncing mechanism becomes dominant over the evaporation process, and offers an elegant way out for the information loss paradox\footnote{See Ref.~\cite{Marolf:2017jkr} for a review on the information loss paradox. It is worth pointing here that large scale quantum gravity effects at the horizon are one possible way out of the information loss paradox. This has led to several proposals, such as the firewall or the fuzzball models. See Ref. \cite{Compere:2019ssx} for a brief overview on this point and Ref. \cite{Rovelli:2019cik} for a recent critical account on the firewall proposal.}, as well as other shortcomings related to the infinite evaporation time and instability of regulars black holes backgrounds discussed so far in the literature \cite{Carballo-Rubio:2018pmi}. More recently, another version of this firework model was proposed, where the black-to-white hole tunneling becomes dominant only at the end of the Hawking evaporation, leading to white hole remnant \cite{Bianchi:2018mml}. The consequences of this alternative picture has been recently discussed in Refs. \cite{Rovelli:2018okm, Rovelli:2018hbk, Martin-Dussaud:2019wqc}. See also Refs. \cite{Baccetti:2018qrp, Baccetti:2016lsb, Baccetti:2019mab, Ho:2019pjr, Ho:2019qiu, Goswami:2005fu} concerning the role of evaporation for collapsing thin shells. Finally, effective black-to-white hole bounces have been recently constructed within the polymer framework \cite{Corichi:2015xia, Olmedo:2017lvt, Ashtekar:2018cay, Bodendorfer:2019cyv, Bodendorfer:2019jay, Bodendorfer:2019nvy, Assanioussi:2019twp}. A crucial difference with, for example, the firework model is that quantum effects are assumed to be dominant only at the Planck scale, replacing the singularity by a transition spacelike hypersurface smoothly connecting a black and a white hole interior effective geometries\footnote{Alternative polymer constructions for black hole interior, taking into account issues related to the covariance of the loop regularization, have been proposed in Refs. \cite{BenAchour:2018khr, Bojowald:2018xxu}. See also Refs. \cite{Aruga:2019dwq, Bojowald:2015zha, Brahma:2014gca} and Refs. \cite{BenAchour:2016brs, BenAchour:2017jof, Bojowald:2019fkv} for extended discussions regarding the problem of covariance in such polymer constructions.}. This picture was recently challenged by another effective Schwarzschild interior model, based on quantum reduced loop quantum gravity \cite{Alesci:2018loi}, that concluded that contrary to current polymer models, the effective dynamics does not predict the formation of an anti-trapped region after the bounce \cite{Alesci:2019pbs}. 

Here, we discuss yet another effective construction to describe a bouncing compact object, keeping as close as possible to the seminal OS model. The reason for this is that the OS model can provide an ideal bridge between bouncing cosmological scenarios and bouncing black holes ones. Indeed, having modeled the interior of the star as a cosmological background, one can replace it by a given bouncing cosmology and try to provide a consistent solution to the matching conditions with the static exterior geometry. There are however a vast number of proposals for both classical and quantum bouncing cosmological scenarios, see Refs.~\cite{Brandenberger:2016vhg, Battefeld:2014uga, Ashtekar:2011ni} for reviews. Therefore, the details of the model will obviously depend on the choice of the bouncing cosmology one started with. In particular, while classical bounces can be realized even if the curvature remain low, quantum bounces are expected to occur only in very high curvature regime. From this perspective, it appears useful to develop a model-independent approach, remaining agnostic on the precise regularization of the singularity at play. As we shall see, important conclusions regarding the energy scale of the bounce and on the allowed range of the parameters can already be extracted using such a strategy. 

In the present work, we revisit the construction of UV-complete gravitational collapse and provide a generalization of the OS model which avoids several limitations affecting previous studies. Following a thin-shell approach, we will glue a static Schwarzschild exterior geometry together with a model-independent bouncing spatially closed FL universe, filled up with homogeneous dust. The time-like thin shell in-between shall be assumed to carry a homogeneous energy and pressure, which encodes part of the effects induced by the corrections beyond GR. We shall also assume that the gluing satisfies the standard Israel-Darmois junction conditions of GR~\cite{Israel:1966rt}. While this assumption can be questioned, it is the simplest choice to construct the model, and it is shared by models such as the black hole firework\footnote{In this model, the authors use the light-like version of the junction conditions discussed in Ref.~\cite{Barrabes:1991ng}.}~\cite{Haggard:2014rza}. Using this setting, the modifications to GR, being motivated classically or of quantum origin, are hidden in the interior bulk as well as in the thin shell dynamics. Solving for the energy and pressure profiles of the thin shell, we will obtain a general model-independent solution describing a UV-complete gravitational collapse. One major challenge of such a construction is to ensure that the conserved quantities associated with the exterior and interior geometries match properly. In the classical OS model, this is ensured by a mass relation. The extension proposed in this work provides two major outcomes:
\begin{itemize}
\item Our first result is the generalization of the OS mass relation, including the effect of the thin shell. This extended mass relation ensures that the mass of the exterior Schwarzschild geometry and the energy of the dust in the interior geometry are indeed constants of motion during the whole process, and it translates into a key constraint on the physical parameters of the model.
\item The second major result of our model-independent construction is that this constraint sets a bound on the energy scale of the bounce. Surprisingly, it implies that if the interior dynamics admits a bounce, it can only occur below, or at most, at the energy scale corresponding to the threshold of horizon formation. Consequently, in our construction, the collapsing star bounces above, or at most at its Schwarzschild radius. This implies that there is no formation of a trapped region contrary to the classical OS scenario. At best, the bouncing star can only form an instantaneous trapping horizon which coincides with the bounce. This constraint is a direct consequence of i) the continuity of the induced metric across the surface of the star and ii) the use of a single horizon vacuum geometry to model the exterior spacetime. This result is completely independent from the choice of modified Friedmann dynamics for the interior geometry as well as the second junction condition involving the discontinuity of the extrinsic curvature. It is worth emphasizing that the condition of continuity of the induced metric is actually shared by all models of higher order gravity, and as such, our result turns out to extend much beyond the present construction.
\end{itemize}
To summarize, the effective theory constructed here provides a simple description of a bouncing compact object which satisfies the conservation of energy during the whole process. This bouncing compact object experiences oscillations between a maximal and a minimal radius, such that no trapped region is ever formed, leading to model-independent realization of an idealized pulsating singularity free compact object\footnote{ See Ref. \cite{Gao:2017ihf} for a similar model.}. Finally, this framework bridges between bouncing cosmology models and bouncing compact objects, and allows on to discuss both classical and quantum bounce scenarii. In a companion paper \cite{paperUS}, we shall present a concrete realization of this model by importing the quantum bounce scenario of spatially closed loop quantum cosmology. This will allow us to implement and test part of the ideas initially introduced in the Planck star model of Ref. \cite{Rovelli:2014cta}.

The work is organized as follows. In Section~\ref{sec1} introduces the model, the effective description of the interior homogeneous region, and derives the effective Friedmann equations we shall work with as well as the energy content of this region. Then, we describe the exterior geometry and introduce the time-like thin shell and describe its energy content. Section~\ref{sec2} presents the model-independent solutions of the junction conditions. We present the first result, namely the generalized mass relation. Then we present the solutions for the energy and pressure profiles of the thin shell, as well as the expression of the lapse as seen by an observer on the surface of the star. It allows us to extract the key constraint which shall be analyzed latter on. Finally, Section~\ref{sec4} studies the implications of this constraint and shows that it sets a bound on the energy scale of the bounce, leading to a marginal horizon-formation as a general conclusion. Section~\ref{sec5} is devoted to a discussion of our construction and the open directions it offers for the future.

\section{Thin shell approach to effective stellar collapse}\label{sec1}
The modeling of stellar collapse follows the seminal work by Oppenheimer and Snyder~\cite{Oppenheimer:1939ue}. It  assumes that the star is described by an ideal ball of dust (i.e. a fluid with vanishing pressure; $p=0$) of constant mass $M$, with a time-dependent density $\rho$, which undergoes a homologous collapse under its own gravity. The full spacetime is then constructed by gluing a time-dependent interior metric to a static exterior metric.

Consider the hypersurface $\Sigma = \mathbb{R}\times \cS^2$ that defines the outer surface of the star and let $g_{\mu\nu}^\pm$ be the interior/exterior metrics, with $(\mu,\nu=0,1,2,3)$ and signature $(-,+,+,+)$. The spacetime embedding of $\Sigma$ is defined by $x^\mu=\bar x^\mu(\sigma^a)$ with $\sigma^a$ intrinsic coordinates. It defines for each of the two metrics an induced metric, $\gamma_{ab}=g_{\mu\nu} e^\mu_a e^\nu_b$ with $e^\mu_a\equiv\partial\bar x^\mu/\partial\sigma^a$, from which one can define the first fundamental form $\gamma_{\mu\nu}\equiv \gamma^{ab}e^\mu_a e^\nu_b$ . It acts as a projector tensor on $\Sigma$ and can be expressed in terms of the unit (spacelike) normal vector to $\Sigma$ as $\gamma_{\mu\nu}=g_{\mu\nu}+n_\mu n_\nu$ such that $n^{\mu}n_{\mu}=+1$. One can then define the extrinsic curvature tensor $K_{\mu\nu}=-\gamma^\alpha_\mu\gamma^\beta_\nu\nabla_\alpha n_\beta$. If $n_\mu$, initially defined only on $\Sigma$, is geodesically extended in the neighborhood of the hypersurface, then $K_{\mu\nu}=-\nabla_\mu n_\nu$. 

The two metrics  $g^{\pm}_{\alpha\beta}$ shall satisfy the Israel-Darmois junction conditions \cite{Israel:1966rt} on the time-like hypersurface $\Sigma$ that defines the outer worldsheet of the star of radius, i.e.
\begin{align}
\label{jun1}
 [\gamma_{\alpha\beta}] & = 0 \\
 \label{jun2}
  [K_{\alpha\beta} - \gamma_{\alpha\beta} K] & = 8 \pi G \sigma_{\alpha\beta}.
\end{align}
The bracket $[X]$ defines the jump of the quantity $X$ at the junction hypersurface, i.e. $[X]=X^+-X^-$. The first equation states that the induced metric is the same whether it is defined from the interior or exterior. $\sigma_{\alpha\beta}$ is  the surface stress-energy tensor of matter fields localized on $\Sigma$.  

We emphasize that using these classical junction conditions is a major assumption of our model. Indeed, when dealing with higher order gravity theories, the junctions conditions turn out to be more involved and for some modified gravity theories, they can prevent the discontinuity of the matter field at the junction hypersurface. This is typically what happens for scalar-tensor theories. However, we adopt the point of view that the set of conditions (\ref{jun1}) and (\ref{jun2}) is the minimal ansatz to build bouncing black hole models. Nevertheless, as mentioned earlier, the first junction condition denoting the continuity of the induced metric remains valid for all known metric theories of gravity.

\subsection{Interior and exterior geometries}
The interior spacetime is assumed to have spatial sections that are homogeneous and isotropic with the topology of a $3$-sphere. Therefore, its metric is of the Friedmann-Lema\^{\i}tre (FL) type
\begin{align}\label{FLmetric}
 \dd s^2_{-} = - \dd\tau^2+  a^2(\tau) R^2_c \left[ \dd \chi^2 + \sin^2{\chi} \dd\Omega^2\right]
\end{align}
where $R_c$ is the constant curvature scale of the spatial sections. $\chi$ is the radial distance from the center, in units of $R_c$, and $a$ the scale factor describing the homogeneous evolution of the star. $\tau$ is the proper time of free-falling observers located at the surface of the sphere defined by
\be
\Sigma: \, \chi=\chi_0.
\ee
By construction, the unit normal vector is radial,  $n_{\mu} \dd x^{\mu} = a(\tau) \dd\chi$. These observers have a $4$-velocity $u^{\mu}$ such that $u_{\mu} n^{\mu} = 0$ given by $u^\mu=(1,0,0,0)$ so that $u_{\mu} \dd x^{\mu} = \dd\tau$ as long as they are comoving. This unit vector remains time-like during the whole evolution, so that $u^{\alpha}u_{\alpha} =-1$. The physical outer radius of the star evolves as
\be
R(\tau) = a(\tau) R_c \sin{\chi_0}.
\ee
The matter inside the star is described by a pressureless fluid with uniform energy density $\rho(\tau)$ so that its stress-energy tensor is $T_{\mu\nu} = \rho (\tau) \; u_{\mu} u_{\nu}$. Its conservation, $\nabla_{\mu} T^{\mu\nu} = 0$, implies
\be\label{CL}
 \rho =  \cE \left(\frac{a}{a_0}\right)^{-3} 
\ee
and we shall set $a_0=1$ without loss of generality. $\cE$ is thus a constant of integration.

The dynamics of the spacetime is derived from the Friedmann equations. In this work, we would like to consider modified Friedman dynamics of a general form. In full generality, we assume they take the form
\begin{align}
\label{HGen}
 \cH^2 & =\left( \frac{8\pi G}{3} \rho - \frac{1}{R_c^2a^2}\right)\left[ 1- \Psi_1(a)\right]\\
  \label{HdotGen}
 \dot{\cH} - \frac{1}{R_c^2a^2}  & = - 4 \pi G  \rho   \left[1 - \Psi_2(a)\right] \\
 \dot{\rho} & = - 3\cH \rho 
\end{align}
where $\Psi_{1,2}(a)$ are dimensionless functions denoting deviations from GR and a dot denotes derivative w.r.t. the time coordinate $\tau$. The Hubble factor $\cH$ is defined as usual as
\begin{eqnarray}
\label{hubbleint}
 \cH(\tau) = \frac{\dot{a}}{a} 
 \end{eqnarray}
When $\Psi_1= \Psi_2 =0$, one recovers the standard Friedmann equations for a closed FL universe. Importantly, note that we require that the classical continuity equation remains satisfied. We stress that by keeping $\Psi_{1,2}(a)$ unspecified at first, we aim to build a general model, such that the model can describe both classical as well as quantum bounces scenarii. The general effective construction can then be applied to any bouncing interior models which can be written in the form of Eqs.~(\ref{HGen}-\ref{HdotGen}). Notice that this is typically the case of the effective dynamics obtained in spatially closed LQC, where the standard regularization of the scalar constraint allows for a classical continuity equation. See Refs. \cite{Ashtekar:2006es, Szulc:2006ep, Corichi:2011pg, Dupuy:2016upu} for details. Before going further, let us already point that the parametrization of the modified Friedman equations introduced above only affects the dynamics of the thin shell, but has no impact on the general constraint we are going to derive in Section~\ref{sec4}, which is purely kinematical. The only assumption we will use is that the correction $\Psi_1$ induces a bounce at some time $\tau_{b}$. Examples of such a correction can be found in Refs. \cite{Ashtekar:2006es, Szulc:2006ep, Corichi:2011pg, Dupuy:2016upu}.
 
Let us now define the initial conditions for the evolution of our system by considering that the star is at equilibrium at $\tau=\tau_0=0$ so that
\be
a(\tau_0) = 1 \qquad \dot{a}(\tau_0) = 0.
\ee
It follows that the maximum radius of the star and the minimum energy density are given by
\be\label{rhomin}
R_{\text{max}} = R_c \sin{\chi_0} \qquad \rho_{\text{min}}  = \cE = \frac{3}{8\pi G R^2_c}.
\ee
This allows us to re-write the Friedmann equations in the more compact form 
\begin{align}
R^2_c\cH^2 & =  \frac{1-a}{a^3} \left( 1- \Psi_1\right) , \\
R^2_c \dot{\cH}  & = \frac{1}{a^3} \left[ a - \frac{3}{2} \left( 1- \Psi_2\right)\right].
\end{align}
Notice that in order for Eq.~(\ref{CL}) to be consistent with the modified Friedmann equations~(\ref{HGen}-\ref{HdotGen}), the two quantum corrections $\Psi_1$ and $\Psi_2$ have to be related through
\be
\Psi_2 = \left( 1- \frac{2a}{3} \right) \Psi_1 - \frac{a \left( 1-a\right)}{3} \frac{\dd \Psi_1}{\dd a}.
\ee
The induced metric $\gamma^{-}_{\alpha\beta}$ and extrinsic curvature $K^{-}_{\alpha\beta}$, as seen from the interior region, are then explicitly given by
\begin{align}
\label{-}
\gamma^{-}_{\mu\nu}\dd x^\mu \dd x^\nu & = - \dd\tau^2+ a^2R^2_c\sin^2{\chi_0} \,\dd\Omega^2, \\
K^{-}_{\mu\nu}\dd x^\mu \dd x^\nu & =   R_c a \sin\chi_0\cos\chi_0\, \dd\Omega^2.
\end{align}
This concludes the presentation of the effective interior spacetime.

Let us now turn to the exterior geometry.
We assume that the exterior metric is well modeled by a classical vacuum spherically symmetric geometry. Thanks to the Birkhoff's theorem, this region is therefore described by the standard vacuum Schwarzschild solution,
\be
\dd s^2_+ = - f(r) \dd t^2 + f^{-1}(r) \dd r^2 + r^2 \dd \Omega^2 
\ee
with
\be
\label{defSchwarz}
 f(r) = 1 - \frac{R_s}{r}
 \ee
where $R_s$ is the Schwarzschild radius
\be\label{e.defRS}
R_s\equiv 2 GM.
\ee
The fact that this geometry possesses only one horizon will have important consequences in the following. Let us furthermore introduce the useful notation
\be
R_s = R_c \sin{\chi_s}
\ee
On the junction hypersurface  $\Sigma$, the exterior coordinates take the parametric form
\be
 t = t_{\ast}(\tau), \qquad r=r_*(\tau)
\ee
where $\tau$ is the proper time of the shell which further coincides with the time coordinate of the interior geometry. Then, the induced metric on $\Sigma$ 
describing the surface of the star is given in Schwarzschild coordinates by
\be
\label{++}
\gamma^{+}_{\mu\nu}\dd x^\mu \dd x^\nu = - A^2[r_*(\tau)]\dd t^2+ r_*^2(\tau) \dd\Omega^2,
\ee
where the new metric function $A[r_*(\tau)]$ of the induced metric is related to the metric function $f(r)$ at $r =r_*(\tau)$ by demanding that the $4$-velocity of the observer co-moving with the sphere of symmetry $\cS^2$ remains a unit time-like vector, i.e $u^{\alpha} u_{\alpha} =-1$. This leads to 
\be
\label{defA}
 A[r_*(\tau)]= \frac{f[r_*(\tau)] }{\sqrt{f[r_*(\tau)] +  \dot{r}^2_{\ast} }}
\ee
The extrinsic curvature of $\Sigma$ induced by the exterior Schwarzschild geometry reads in term of the (FL) coordinates system $(\tau, \theta, \phi)$
\be
\label{+++}
K^{+}_{\mu\nu}\dd x^\mu \dd x^\nu =  - \frac{2A[r_{\ast}(\tau)]}{f[r_{\ast}(\tau)]} \left[ \ddot{r}_{\ast} + \frac{1}{2} f'[r_{\ast}(\tau)] \right] \dd \tau^2 +\frac{r_*f[r_*(\tau)]}{A[r_*(\tau)]}\dd\Omega^2.
\ee
Let us now describe the energy content of the shell.

\subsection{Introducing a surrounding thin shell}

In full generality, the matching hypersurface can enjoy a non vanishing surface stress-energy tensor, that could arise from the tension of $\Sigma$. As we will recall below, it vanishes in the original OS model~\cite{Oppenheimer:1939ue}. We assume that the thin shell is described by a perfect fluid with surface energy density $\sigma$ and pressure $\Pi$ so that the surface stress tensor $\sigma_{\alpha\beta}$ reads 
\be
\sigma_{\alpha\beta} = \epsilon \sigma u_{\alpha} u_{\beta} + \Pi \left( h_{\alpha\beta} + u_{\alpha} u_{\beta}\right),
\ee
that is $\sigma^\alpha_\beta={\rm diag}(\epsilon \sigma, \Pi, \Pi)$. In this work, the thin shell is time-like (as it should be for any type of physical matter), which implies $\epsilon =-1$. 

\section{Junction conditions and generalized mass relation}\label{sec2}
We can now write down the junction conditions (\ref{jun1}-\ref{jun2}) using Eq.~(\ref{-}) and Eqs. (\ref{++}-\ref{+++}). The continuity of the induced metric~(\ref{jun1}) implies that
\begin{eqnarray}
\label{fjun1}
 r_*(\tau) &=& R_c a(\tau)\sin\chi_0 = R(\tau), \label{e.rstar}\\
 \label{fjun2}
\frac{\dd t_{\ast}}{\dd\tau}&=& A^{-1}(\tau), \label{ddt}
\end{eqnarray}
which give the evolution of the surface of the star in the Schwarzschild region and the relation between the times of comoving observers in the two regions. Furthermore, they imply that the evolution of the radius of the outer surface of the star is given by
\be
\label{e.rdot}
\dot{r}_{\ast} =  r_*(\tau) \cH(\tau) .
\ee
where $\cH(\tau)$ is given by (\ref{hubbleint}).
Finally, the jump in the extrinsic curvature, encoded in Eq.~(\ref{jun2}) gives the expressions of the surface stress-energy tensor.
\begin{align}
\label{excurv1}
8 \pi G \; \sigma (\tau) &= \frac{1}{r_{\ast}(\tau)} \left[ \frac{f[r_{\ast}(\tau)]}{A[r_{\ast}(\tau)]} - \cos{\chi_0}\right], \\
\label{excurv2}
8 \pi G\;  \Pi (\tau)& = - \frac{2A[r_{\ast}(\tau)]}{f[r_{\ast}(\tau)]} \left[ \ddot{r}_{\ast} + \frac{1}{2} f'[r_{\ast}(\tau)] \right].
\end{align}
The goal now is to solve for the profile of the energy $\sigma$ and the pressure $\Pi$ using the effective Friedmann equations and the two constants of motion associated to the exterior and interior geometries. We will see that non-trivial consistency conditions restrict the range of the free parameters of the model in order to have physically acceptable profiles for $\left( \sigma, \Pi\right)$.

\subsection{Generalized mass relation}

Consider first the equation (\ref{excurv1}). It can be easily recast as
\begin{align}
\cos{\chi_0} + 8 \pi G r_{\ast}(\tau) \sigma = \sqrt{f[r_*(\tau)] + R^2_c a^2(\tau)\sin^2{\chi_0} \cH^2(\tau)}
\end{align}
where we have used Eq.~(\ref{e.rdot}) to express $\dot r_*/A$. After some algebra and upon replacing the Hubble factor $\cH$ using the general first Friedmann equation (\ref{HGen}), one obtains finally one of the master equations relating the constant mass of the exterior Schwarzschild geometry $M$, the energy density $\rho(t)$ of the perfect fluid and the surface stress energy $\sigma(t)$ of the thin shell. This relation reads
\begin{align}
\label{Master-Equation}
M = \frac{4\pi }{3} \rho r^3_{\ast} -  \frac{r^3_{\ast}}{2G R^2_c} \left[ \Sigma^2 + \frac{2 R_c \cos{\chi_0 }}{r_{\ast}} \Sigma +  \frac{ 1-a}{a^3} \Psi_1 \right],
\end{align}
once we introduce the dimensionless quantity  $\Sigma= 8\pi G R_c \sigma$.

At this point, few remarks are in order. The first term of the r.h.s. is a conserved quantity associated with the perfect fluid related to the constant ${\cal E}$ defined in Eq.~(\ref{CL})
\be
\frac{4\pi}{3} \rho r^3_{\ast} = \frac{4\pi}{3} \cE R^3_c \sin^3{\chi_0},
\ee
In the limit $\Psi_1 = 0$ (pure GR) and $\Sigma=0$ (no shell), the above relation reduces to the standard OS mass relation,
\be
\label{OSstandard}
M \big{|}_{\sigma = \Psi_1 = 0}  = \frac{4\pi}{3} \rho r^3_{\ast}.
\ee
However, when a thin shell is present, the general relation is given by Eq.~(\ref{Master-Equation}). This leads to a second order polynomial equation, that determines the associated profile for the surface energy density of the thin shell. 

\subsection{The thin shell energy profile}

We turn now to solving for the energy profile for $\Sigma$ from the equation
\be
\label{sop}
\Sigma^2 + \frac{2 \cot{\chi_0}}{a} \Sigma  + \frac{1}{a^3} \left[ \frac{\sin{\chi_s}}{ \sin^3{\chi_0}} - 1 +  \left( 1-a\right) \Psi_1 \right]=0 .
\ee
A consistent solution for the profile of $\Sigma$ is obtained if, and only if, the determinant of this second order equation is positive, i.e.
\begin{align}
\label{det}
\Delta_{\Sigma} & = \frac{4}{a^2} \left\{ \cot^2{\chi_0} - \frac{1}{a} \left[ \frac{\sin{\chi_s}}{\sin^3{\chi_0}} - 1 + \left( 1-a\right) \Psi_1 \right]\right\} > 0 .
\end{align}
This provides a constraint on the range of the parameters of our model. When this condition is satisfied, the junction conditions admit real solutions for the energy of the thin shell, given by
\be
\label{sigmaprof}
\Sigma_{\pm}  = \frac{1}{aR_c} \left\{  - \cot{\chi_0} \pm \sqrt{\Delta_{\Sigma}} \right\}.
\ee
It will be convenient to single out the standard GR part thanks to the splitting $\Delta_{\Sigma}  =  \Delta^{\text{GR}}_{\Sigma}  + \Delta^{\text{QG}}_{\Sigma}$ with
\begin{align}
\Delta^{\text{GR}}_{\Sigma}  =  \frac{4}{a^2R^2_c} \left[ \cot^2{\chi_0} - \frac{1}{a} \left( \frac{\sin{\chi_s}}{ \sin^3{\chi_0}} - 1 \right)\right],  \qquad
 \Delta^{\text{QG}}_{\Sigma}  = - \frac{4}{R^2_c}   \frac{1-a}{a^3} \Psi_1,
\end{align}
so that $\Sigma_{\pm}= \Sigma_{\pm}^{\text{GR}} + \Sigma_{\pm}^{\text{QG}}$ with
\begin{align}
\label{sigmaRG}
\Sigma_{\pm}^{\text{GR}} & = \frac{1}{aR_c} \left\{  - \cot{\chi_0} \pm \sqrt{\Delta^{\text{GR}}_{\Sigma}} \right\}.
\end{align}

Let us now discuss some useful limits. In the infrared limit, namely $a \rightarrow 1$, the energy of the thin shell has the non-vanishing value
such that the effective quantum correction $\Psi_1$ does not affect the classical regime, whatever the form of the UV completion, thanks to the term $(1-a)$. Let us further assume that in the pre-collapse configuration, the thin shell does not violate the weak energy condition, namely that $\Sigma_{\pm} (\tau=0, \chi_0, \chi_s) \geqslant 0$. This is true if we restrict the parameters of the model $\left( \chi_0, \chi_s\right)$ to satisfy
\be
\sin{\chi_s} \leqslant \sin^3{\chi_0},
\ee
the equality corresponding to the OS configuration when $\Psi_1=0$.

\subsection{The lapse at the surface of the star}

The lapse function shall play a crucial role when discussing the energy scale of the bounce in the next section.  Using its expression (\ref{defA}), and invoking expression (\ref{e.rdot}), one obtains 
\begin{align}
\label{AA11}
A^2[r_*(\tau)]  = \frac{f^2[r_*(\tau)]}{f[r_*(\tau)] + r^2_{\ast} \cH^2} .
\end{align}
Moreover, the lapse can be also related to the determinant (\ref{det}) of the  mass relation as follows. Combining the first junction condition (\ref{excurv1}) and the expression for the energy of the thin shell (\ref{sigmaprof}), one obtains that
\be
\label{Aaa}
A^2[r_*(\tau)] = \frac{4f^2[r_*(\tau)]}{a^2 \sin^2{\chi_0} \Delta_{\Sigma}},
\ee
We shall see in Section~\ref{sec4} that the expression for the lapse will allow us to extract very generic constraints on the allowed energy scale of the bounce.

\subsection{The thin shell pressure profile}

The pressure profile can then be determined from the junction condition~(\ref{excurv2}). Since
\begin{align}
\ddot{r}_{\ast} &   =  r_{\ast} \left( \cH^2 + \frac{\dd \cH}{\dd\tau}\right)  ,
\end{align} 
the pressure profile takes the form
\begin{align}
\label{pressprof}
& 8 \pi G \Pi = - \frac{2A[r_*(\tau)]}{f[r_*(\tau)]} \left[   r_{\ast}  \left( \cH^2 + \frac{\dd \cH}{\dd\tau}\right)  +  \frac{R_s}{2r^2_{\ast}} \right]
\end{align}
where the function $A$ is given by Eq.~(\ref{Aaa}). 

This concludes the construction of the effective theory of our stellar collapse model beyond GR. Eqs. (\ref{sigmaprof}) and~(\ref{pressprof}) provide the profiles of the surface pressure and surface energy of the thin shell in the general case where the interior of the star has an effective modified dynamics modeled by Eqs.~(\ref{HGen}) and (\ref{HdotGen}). Notice that the mass of the star, given by Eq.~(\ref{Master-Equation}) remains conserved during the whole evolution. Before presenting the main outcome of this model, let us make two more remarks. First, we observe that the presence of the thin shell implies, through the generalized mass relation (\ref{Master-Equation}), that the Schwarzschild mass $M$ and the constant of motion associated to the dust $\cE  \sim \rho(\tau) R^3(\tau)$ do not match anymore as in the classical case. This implies that these two constants can now be chosen arbitrarily, enlarging the parameters space describing the new compact object. Moreover, the interior being modeled by a bouncing cosmology, one expects that if one introduces a small inhomogeneity prior to collapse, it would grow during the cycles of contraction and expansion, to finally lead to a strong deviation from spherical symmetry, breaking the idealized effective description discussed here. Quantifying precisely the number of cycles during which the effective description remains valid would require to pick up a given model for the interior, through a concrete choice of $\Psi_1$.  

\section{UV-completeness and constraints on the energy scale of the bounce}\label{sec4}
So far, we have obtained a general solution of the thin shell satisfying the junction conditions~(\ref{jun1}-\ref{jun2}) that allows for the construction of a spacetime in which the interior solution enjoys a modification of GR while the exterior is still described by a vacuum Schwarzschild spacetime. This solution is characterized by the generalized mass relation (\ref{Master-Equation}) and the associated profiles for the energy and pressure of the thin shell given by Eqs.~(\ref{sigmaprof}) and~(\ref{pressprof}). 

Let us now discuss the generic constraint imposed by the consistency of such a gluing on the energy scale of the bounce. The crucial equation is the expression of the lapse function at the surface of the star, which we reproduce here for the ease of the reader:
\begin{align}
\label{AA11new}
A^2[r_*(\tau)]  = \frac{f^2[r_*(\tau)]}{f[r_*(\tau)] + r^2_{\ast} \cH^2},
\end{align}
with  $f[r_*(\tau)]= 1 - R_s/r_*(\tau)$ is the usual Schwarzschild factor~\eqref{defSchwarz}.  First, note that the induced metric on the thin shell is continuous, thanks to the first of the Israel-Darmois junction conditions~(\ref{jun1}) and, more importantly, that this is completely independent of the physical properties of the thin shell. This is an extremely general condition, which actually goes beyond GR. One might correctly point out that the Israel-Darmois junction conditions for thin shell dynamics crucially depend on Einstein's equations and thus our formalism borrows this from GR. Nevertheless, let us emphasize that changes to the junction conditions from modified gravity theories beyond GR is expected to show up in the equation for the extrinsic curvature. The derivation of Eq.~\eqref{AA11new} above only requires the continuity of the induced metric which is expected to be satisfied in all known higher order modified gravity theories such as $f(R)$ or scalar-tensor theories. Physically, it seems rather reasonable to assume that the induced metric should not depend on which side of the thin shell one derives it from unless there are some violent, and perhaps undesirable, departures from covariance.

Having assumed the continuity of the induced metric, we then only have to use the definition~\eqref{defA} of the lapse function  to get Eq.~\eqref{AA11new}. Since we are considering models which exhibit a bounce, thereby avoiding the classical singularity, the effective Friedmann equation must satisfy $\cH (\tau_b) = 0$, at the time of the bounce $\tau_b$, where $\cH$ is the usual Hubble parameter. Thus, provided a bounce occurs, the lapse function at the bounce reads 
\be
A^2(\tau_b) =   1-\frac{R_s}{r_{\ast}(\tau_b)},
\ee
which remains well-defined provided the radius of the star at the bounce satisfies $r_{\ast}(\tau_b)  \geqslant R_s$. Since this is our main observation, let us summarize that as
\be
 \qquad  \cH\left(\tau_{\text{b}}\right)=0 \qquad \Rightarrow \qquad r_{\ast} \left(\tau_{\text{b}}\right) \geqslant R_s.
\ee 
Consequently, the existence of a bounce implies that the contracting star can at most reach its Schwarzschild radius. The bounce \textit{must} occur above or at the threshold of horizon formation, setting an upper bound for the energy scale of the bounce mechanism. Notice that the use of the Schwarzschild geometry plays also a crucial role in this result, in that it exhibits one single horizon. It is worth keeping in mind that with another exterior geometry, with possibly multiple horizons, this inequality might lead to a different outcome, as it would involve also an additional inner horizon structure. This generalization shall be studied in upcoming work \cite{futureUS}. 

Now, in order to show that the above inequality is \textit{not} a gauge artifact of working with a singular coordinates choice for the Schwarzschild metric, we present an alternative computation of our key relation (\ref{AA11new}) in the Eddington-Finkelstein coordinates. This would also allow the reader to see the simplicity of our argument. In these coordinates, the metric is explicitly non-singular at the Schwarzschild radius and therefore $r=2M$ does not present itself as a special point. For concreteness, we work with the ingoing Eddington-Finkelstein gauge, in which the metric takes the form
\begin{eqnarray}
\dd s^2_+ = - f(r) \dd v^2 + 2 \dd v \dd r + r^2 \dd \Omega^2. 
\end{eqnarray}
With this form of the external metric,  the induced  metric on the hypersurface  $\Sigma: \, r=r_*(v)$ is
\begin{eqnarray}
\label{EF++}
	\gamma^{+}_{\mu\nu}\dd x^\mu \dd x^\nu = - N^2[r_*(v)]\dd v^2+ r_*^2(v) \dd\Omega^2\,.
\end{eqnarray}
The lapse function $N$ relates the `time' coordinate of the ingoing Eddington-Finkelstein metric $v$ to the proper time of the interior FL metric $\tau$.  Expressing the lapse function in terms of the metric coefficient $f(r)$, one gets
\begin{eqnarray}
	\label{defN}
	N[r_*(v)] = \sqrt{f[r_*(v)] - 2\dot{r}_*}\,,
\end{eqnarray}
where the dot now refers to a derivative with respect to $v$. By demanding the continuity of the induced metric on $\Sigma$ at the point $r=r_*$, the Israel-Darmois junction condition~(\ref{jun1}) for the induced metric takes the form
\begin{eqnarray}
	r_* &=& R_c \sin\chi_0 \,a, \label{EFe.rstar}\\
	 N &=& \frac{\dd \tau}{\dd v}. \label{EFddv}
\end{eqnarray}
This is easy to derive by comparing the two expressions~\eqref{EF++} and~(\ref{-}). Hence, the evolution of the surface of the star is given by
\begin{eqnarray}
\label{EFe.rdot}
\dot r_*(v) =  r_*(v) N[r_*(v)] \mathcal{H}.
\end{eqnarray}
From Eqs.~\eqref{EFe.rdot} and~\eqref{defN}, it is easy to show that
\begin{eqnarray}
	N = -r_* \mathcal{H} \pm \sqrt{r_*^2\mathcal{H}^2 + f}\,.
\end{eqnarray}
For a bouncing model, once again, we require that $\mathcal{H}(\tau_b) = 0$, which leads to the same conclusion, i.e. that the bounce must occur outside, or at best at, the Schwarzschild radius. The important point is that this conclusion holds even when we do not have an apparent (coordinate) singularity in our choice of the eternal metric. Rather, the important condition is that a smooth matching of the induced metric  is performed on a time-like surface. 

Let us repeat the crucial point. We never needed to use the second junction condition, involving the extrinsic curvature and the physical properties of thin-shell, to reach this conclusion. Indeed, this latter junction condition~(\ref{jun2}) fully determines the stress-energy tensor of the thin shell and depends on the details of the quantum, or effective, gravity theory. This can also be rephrased by saying  that albeit we require that the effective Friedmann equation allows for a bouncing solution (i.e.  $\cH=0$ et some time) we do not need any further details on the exact form of the Friedmann equations for our argument. Such an exact form would, of course, depend on the generalization to GR provided by the effective UV-theory, but our conclusions remain completely independent of it. Hence, this constraint is purely kinematical\footnote{While the constraint on the bounce is derived from the first junction condition involving the continuity of the induced metric across the time-like thin shell, as as such is a kinematical statement, is has an interesting dynamical interpretation. Indeed, the constraint can also be understood from the expression (\ref{Aaa}) for the lapse function $A(\tau)$ in term of the determinant $\Delta(\tau)$. The condition $\Delta(\tau) \geqslant0 $ ensures the existence of real solution for the energy profile of the thin shell, which coincides with the requirement that $A^2 \geqslant 0$. This shows the consistency between the kinematical constraint and its dynamical realization.}. In the end, the bouncing compact object described in the present model is constrained by the following conditions
\be
\label{con22}
a_{\text{min}} \geqslant \frac{R_s}{R_{\text{max}}} \;, \qquad \text{and} \qquad R_s \leqslant  R_{\text{max}} \leqslant R_c
\ee
where $a_{\text{min}}$ is the minimal allowed value of the scale factor, corresponding to the bouncing point. The first condition is rather intriguing as for macroscopic objects, it implies that the modifications to GR can be dominant even in regime of low curvature. The possibility of such a counter-intuitive behavior in the present model can be explained as follows.

Let us first point out that this construction being model-independent, we do not make the difference between classical versus quantum bounces. However, these two types of bounces are quite different in their physical origin. Indeed, it is worth pointing that for classical bounces, such as the ones obtained in modified gravity theories, they can occur at low curvatures (see Refs \cite{Easson:2011zy, Qiu:2011cy, Cai:2012va, Li:2014era, Qiu:2015nha, Kolevatov:2017voe, Ijjas:2016tpn, Dobre:2017pnt} for details). On the contrary, it is expected that quantum bounces, which originate from quantum gravity proposals, will occur only in the very high curvature regime. Consequently, for the classical bounce scenario, the constraint (\ref{con22}) can be easily realized, but for quantum bounce, demanding additionally that $a_{\text{min}}$ coincides with high curvature regime severely restricts the domain of application of the present construction. As an example of this, we construct, in a companion paper \cite{paperUS}, a concrete realization of such bouncing objects where the bounce is assumed to be of quantum origin due to LQC effects. Demanding that the UV cut-off of the effective dynamics be of order of the Planck length, it is shown that the consistency of the model excludes macroscopic stellar objects but allows one to model only Planckian relics. For such small size objects, the Schwarzschild radius is pushed to very small value, and the curvature at horizon scale can be huge\footnote{And, yet, it is important to remember there are quantum models which predict a bounce near Planckian energy densities, whereas $a_{\text{min}}$ remains well above Planck length.}. This reconciles our no-go result with the intuitive expectation that quantum bounces should occur only in a high curvature regime. However, the bigger point which our result implies is that  one would have to contend with the ``leaking out'' of quantum effects from the interior -- and that the underlying quantum theory playing a big role at (or, outside) the horizon -- \textit{unless} one is willing to add extra structure to the quantum vacuum black hole geometry such as an inner horizon. Indeed, with hindsight, we expect that quantum effects within the horizon should lead to such additional structures for consistent black-to-white hole bounces.


This gives us the proper segue to remind the readers about our assumptions of  i) spherical symmetry and ii) a vacuum exterior geometry for the star. Then, if classicality is assumed for the exterior geometry, Birkhoff's theorem implies that the geometry out of the Planck star is static and described by the Schwarzschild solution with the exterior geometry exhibiting only a single horizon. If one relaxes the assumption of a classical vacuum geometry to model the exterior of the star, and indeed one should expect so near the surface, one could use some \textit{ad hoc} regular black hole metric (or, preferably, one derived form some quantum gravity theory) with possibly multiple horizons. In that case, the key inequality would involve the horizon structure of the exterior geometry, and the presence of possible inner horizon would allow the formation of a trapped region. This can potentially lead to consistent models of black-to-white hole bounce, with the bounce occurring far within the (outer) horizon, and, therefore at high curvatures. The natural next step of this work is therefore to extend this model to a quantum vacuum geometry for the exterior of the star and check precisely how the current scenario is modified. 

To reiterate our main lesson regarding the construction of quantum black-to-white hole bounces, we find that  either one accepts that bounces happen at low curvatures, which goes against all standard intuitions, or accepts that vacuum LQG models such as \cite{Corichi:2015xia, Olmedo:2017lvt, Ashtekar:2018cay, Bodendorfer:2019cyv, Bodendorfer:2019nvy, Bodendorfer:2019jay} are currently missing a crucial ingredient which might be the formation of an inner horizon during the collapse, or similar additional structures. Only then one could consistently keep quantum gravity effects confined to the deep interior, in a high curvature regime, while working with such a dynamical model of matter collapse (as opposed to that of an eternal balck hole).

\section{Discussion}\label{sec5}
This article provides a UV completion of the seminal OS collapse which allows one to discuss generic features of an effective bouncing compact object. Our model-independent approach relies on the thin shell formalism to glue a quantum bouncing Friedmann-Lema\^{\i}tre closed universe to an exterior vacuum Schwarzschild geometry. The mechanism responsible of the UV completion is encoded in the corrections to the Friedmann equations, i.e. $\Psi_1$, which are kept unspecified. The gluing is assumed to be captured by the classical Israel-Darmois junction conditions of GR, which constitutes a major simplifying assumption. The first result presented in Section~\ref{sec2} is an explicit model-independent solution of the junction conditions, which describes a bouncing star. The corrections turn out to affect both the interior geometry as well as the dynamics of the thin shell. Let us emphasize the key points of this construction.

Consistency of the solution is encoded in the mass relation (\ref{Master-Equation}), which generalizes the standard classical OS mass relation (\ref{OSstandard}). This extension allows one to properly match the conserved quantities associated to the exterior and interior regions, namely the ADM mass of the Schwarzschild geometry and the energy of the dust filling up the interior of the star. Upon solving this key constraint, one obtains the general solution for the energy and pressure of the thin shell. During the whole evolution, these surface quantities adapt such that the mass relation is always satisfied. One major advantage of this model is therefore to include in a consistent way the role of the matter as driving the collapse, a step which has been largely left unexplored in most of bouncing black hole models focusing on vacuum interior geometries \cite{Corichi:2015xia, Olmedo:2017lvt, Ashtekar:2018cay, Bodendorfer:2019cyv}. Moreover, compared to the black-hole firework model \cite{Haggard:2014rza}, where matter is encoded in null shells, our model provides an alternative way to include the matter sector through a collapsing dust field. This is a crucial difference as it allows the model to work with a time-like junction hypersurface, which in turn is one of the key assumption to derived the no-go result regarding the formation of a trapped region. Finally, the main advantage of our model is to provide model-independent conclusions regarding the UV-completion of the OS collapse.

Indeed, as we have explained in details in Section~\ref{sec4}, a major outcome of this construction is that the bouncing compact object never forms a trapped region and bounces below, or at most at, the energy scale corresponding to the threshold of horizon formation. Consequently, there can be no black hole formation as opposed to the classical Oppenheimer-Snyder model, as only an instantaneous (w.r.t. the thin shell frame) event horizon can form at the bounce. This surprising result descends solely from i) the demanding continuity of the induced metric and ii) a (classical) vacuum Schwarzschild geometry outside the star. In particular, the second junction condition involving the extrinsic curvature, and which is the only one typically modified in higher order gravity theories, plays no role what so ever in deriving this result. A crucial ingredient in deriving this result is that the gluing is performed across a time-like thin shell. This directly leads to the expression (\ref{AA11new}) for the lapse at the surface of the star. Note that this expression, and the resulting constraint, are both absent if one uses instead a null-shell to glue the exterior and interior geometries, as it is done for example in Ref \cite{Haggard:2014rza}. This explains the crucial difference between the present model and previous constructions based on a collapsing null-shell. 


Moreover, the present construction provides an example where the scale at which quantum gravity effects become dominant is not fixed \textit{a priori}, but is imposed by the internal consistency of the construction. On first glance, our results do suggest that the bounce must take place outside (or, at best, at) horizon scales. Although this can be achieved for models based on classical bounces, this seems to be rather restrictive for regular black-to-white hole models originating from quantum gravity approaches since one does not expect to have huge quantum effects at such low curvatures. However, see Refs.~\cite{Carballo-Rubio:2019nel, Carballo-Rubio:2019fnb} for some generic arguments in this direction. Moreover, from typical quantum gravity arguments, it has been proposed in Refs. \cite{Barcelo:2014npa, Barcelo:2014cla} that if one adopts a smoothing procedure to halt classical gravitational collapse due to the appearance of some minimum radius, a shock wave is generated which propagates outwards. Nevertheless, it was already noticed in Ref. \cite{Brahma:2018cgr} that the radial extent of this curvature wave would be felt even outside the event horizon although its origin lies deep inside. Similar conclusions have also been shown to hold for the ``firework'' model discussed in Ref. \cite{Haggard:2014rza} and its variations in Ref. \cite{DeLorenzo:2015gtx}, in that the effects of a black-to-white-hole transition would be felt outside the horizon \cite{Brahma:2018cgr}. Provided one works with the same simplifying assumptions regarding the exterior geometry, the present model-independent construction points in the same direction, showing that quantum gravity effects can kick in already above or at horizon scale, going against all intuitions. 

Nevertheless, as has been pointed out, this is not the interpretation we wish to forward from our findings. Indeed, we do not suggest that quantum effects necessarily extend outside the horizon (at very low curvatures), but rather use our consistency conditions to point towards the inadequacy of current quantum models of vacuum Schwarzschild spacetimes \cite{Corichi:2015xia, Olmedo:2017lvt, Ashtekar:2018cay, Bodendorfer:2019cyv, Bodendorfer:2019nvy, Bodendorfer:2019jay}. The physical requirement coming from the OS collapse of matching the induced metric across the (time-like) surface of the star seems to point towards additional structures in the quantum black hole interiors, such as an inner horizon. In other words, effective black-to-white hole models, which are based on quantum bounces and not on classical ones coming from modified gravity, need to satisfy an extra physical requirement when considering matter collapses (as opposed to eternal black holes) which comes in the form of requiring an inner horizon. The appearance of such structures would ensure that consistent quantum bounces can still be confined to the deep interior while satisfying the consistency conditions unveiled in our work. In developing such a refined model which would keep the quantum effects confined weel within the outer horizon, we shall also relax our assumption of classical Schwarzschild geometry near the exterior of the surface of the star. On the other hand, if we want to insist on a classical Schwarzschild spacetime outside the star, with only one horizon, then this model tells us that consistent bounces of quantum origin can only happen for Planckian relics, see the companion paper \cite{paperUS}. 

Finally, let us point out that a necessary requirement for our formalism to go through is, obviously, that the `effective' approach is valid in the sense of the semiclassical approximation. Typically, for effective models, it is implied that one can work with the expectation values of quantum operators taken about some well-defined semiclassical states, while ignoring fluctuations and other higher moments. Indeed, it has been shown that the effective equations can be trusted in the context of cosmology \cite{Rovelli:2013zaa}, and more generally for quantum field theories \cite{Bojowald:2015fla}, in the semiclassical approximation. As long as the mass of the collapsing objects is much larger than the Planck mass, we would expect similar effective equations to be valid for black-to-white hole bouncing scenarios as well. (In particular, notice that order of magnitude for the mass of the Planckian relics considered in \cite{paperUS} would still be much bigger than the Plank mass so that these effective equations remain valid for them). Once again, for any consistent theory, we expect quantum effects to be confined in the deep interior such that, as long as we far away from Planckian length scales, we can expect our degrees of freedom to be well-described by these effective equations. This is another way of arguing that contrary to appearances, our results do not really imply having large quantum effects outside the horizon but rather strongly hint at the existence of quantum black hole models which would allow for an inner horizon so that these quantum effects can be kept confined to the high curvature regimes. In short, it is not the failure of the effective equations which can circumvent our no-go results but rather the existence of additional structure in the quantum black hole constructions which must do so.

\section*{Acknowledgements}
The work of JBA was supported by Japan Society for the Promotion of Science Grants-in-Aid for Scientific Research No. 17H02890.
SB is supported in part by funds from NSERC, from the Canada Research Chair program, by a McGill Space Institute fellowship and by a generous gift from John Greig. The authors thank Norbert Bodendorfer and Edward Wilson-Ewing for helpful suggestions on a previous version of this draft.


\begin{thebibliography}{ab}

  \bibitem{Giddings:1992hh} 
  S.~B.~Giddings,
  ``{Black holes and massive remnants,}''
  Phys.\ Rev.\ D {\bf 46}, 1347 (1992)
  \href{http://arXiv.org/abs/9203059}{{\texttt{arXiv:9203059}}}
  

\bibitem{Giddings:2016btb} 
  S.~B.~Giddings and D.~Psaltis,
  ``{Event Horizon Telescope Observations as Probes for Quantum Structure of Astrophysical Black Holes,}''
  Phys.\ Rev.\ D {\bf 97}, no. 8, 084035 (2018)
  \href{http://arXiv.org/abs/1606.07814}{{\texttt{arXiv:1606.07814}}}
 
  
  \bibitem{Giddings:2019jwy} 
  S.~B.~Giddings,
  ``{Searching for quantum black hole structure with the Event Horizon Telescope,}''
  Universe {\bf 5}, no. 9, 201 (2019)
  \href{http://arXiv.org/abs/1904.05287}{{\texttt{arXiv:1904.05287}}}
  
  \bibitem{Almheiri:2012rt} 
  A.~Almheiri, D.~Marolf, J.~Polchinski and J.~Sully,
  ``{Black Holes: Complementarity or Firewalls?,}''
  JHEP {\bf 1302}, 062 (2013)
  \href{http://arXiv.org/abs/1207.3123}{{\texttt{arXiv:1207.3123}}}
  
  \bibitem{Almheiri:2013hfa} 
  A.~Almheiri, D.~Marolf, J.~Polchinski, D.~Stanford and J.~Sully,
  ``{An Apologia for Firewalls,}''
  JHEP {\bf 1309}, 018 (2013)
  \href{http://arXiv.org/abs/1304.6483}{{\texttt{arXiv:1304.6483}}}
  
  \bibitem{Mazur:2004fk} 
  P.~O.~Mazur and E.~Mottola,
  ``{Gravitational vacuum condensate stars,}''
  Proc.\ Nat.\ Acad.\ Sci.\  {\bf 101}, 9545 (2004)
  \href{http://arXiv.org/abs/0407075}{{\texttt{arXiv:0407075}}}
  
  \bibitem{Mathur:2008nj} 
  S.~D.~Mathur,
  ``{Fuzzballs and the information paradox: A Summary and conjectures,}''
  \href{http://arXiv.org/abs/0810.4525}{{\texttt{arXiv:0810.4525}}}
 
  
  \bibitem{Maldacena:2013xja} 
  J.~Maldacena and L.~Susskind,
  ``{Cool horizons for entangled black holes,}''
  Fortsch.\ Phys.\  {\bf 61}, 781 (2013)
  \href{http://arXiv.org/abs/1306.0533}{{\texttt{arXiv:1306.0533}}}

\bibitem{Barrau:2019swg} 
  A.~Barrau, K.~Martineau, J.~Martinon and F.~Moulin,
  ``{Quasinormal modes of black holes in a toy-model for cumulative quantum gravity,}''
  Phys.\ Lett.\ B {\bf 795}, 346 (2019)
  \href{http://arXiv.org/abs/1906.00603}{{\texttt{arXiv:1906.00603}}}

\bibitem{Carballo-Rubio:2018jzw} 
R.~Carballo-Rubio, F.~Di~Filippo, S.~Liberati, and M.~Visser,
  ``{Phenomenological aspects of black holes beyond general relativity},''
  Phys. Rev. {\bf D98} (2018), no.~12, 124009,
\href{http://arXiv.org/abs/1809.08238}{{\texttt{arXiv:1809.08238}}}.

\bibitem{Malafarina:2017csn}
D.~Malafarina, ``{Classical collapse to black holes and quantum bounces: A
  review},'' Universe {\bf 3} (2017), no.~2, 48,
\href{http://arXiv.org/abs/1703.04138}{{\texttt{arXiv:1703.04138}}}.

\bibitem{Barrau:2018rts} 
  A.~Barrau, K.~Martineau and F.~Moulin,
  ``{A status report on the phenomenology of black holes in loop quantum gravity: Evaporation, tunneling to white holes, dark matter and gravitational waves,}''
  Universe {\bf 4}, no. 10, 102 (2018)
  \href{http://arXiv.org/abs/1808.08857}{{\texttt{arXiv:1808.08857}}}.

\bibitem{Oppenheimer:1939ue}
J.~R. Oppenheimer and H.~Snyder, ``{On Continued gravitational contraction},''
  Phys. Rev. {\bf 56} (1939)
455--459.


\bibitem{Bojowald:2015iga}
M.~Bojowald, ``{Quantum cosmology: a review},'' Rept. Prog. Phys. {\bf 78}
  (2015) 023901,
\href{http://arXiv.org/abs/1501.04899}{{\texttt{arXiv:1501.04899}}}.

\bibitem{Frolov:1979tu}
V.~P. Frolov and G.~A. Vilkovisky, ``{Quantum gravity removes classical singularities and shortens the life of black holes},'' in {\em {The Second
  Marcel Grossmann Meeting on the Recent Developments of General Relativity (In
  Honor of Albert Einstein) Trieste, Italy, July 5-11, 1979}}, p.~0455.
\newblock
1979.
\newblock

\bibitem{Frolov:1981mz}
V.~P. Frolov and G.~A. Vilkovisky, ``{Spherically Symmetric Collapse in Quantum
  Gravity},'' Phys. Lett. {\bf 106B} (1981)
307--313.

\bibitem{Hajicek:1992mx}
P.~Hajicek, ``{Quantum mechanics of gravitational collapse},'' Commun. Math.
  Phys. {\bf 150} (1992)
545--559.

\bibitem{Hajicek:2001yd}
P.~Hajicek and C.~Kiefer, ``{Singularity avoidance by collapsing shells in
  quantum gravity},'' Int. J. Mod. Phys. {\bf D10} (2001) 775--780,
\href{http://arXiv.org/abs/gr-qc/0107102}{{\texttt{arXiv:gr-qc/0107102}}}.

\bibitem{Schmitz:2019jct} 
  T.~Schmitz,
  ``{Towards a quantum Oppenheimer-Snyder model,}''
  Phys.\ Rev.\ D {\bf 101}, no. 2, 026016 (2020)
  \href{http://arXiv.org/abs/gr-qc/1912.08175}{{\texttt{arXiv:gr-qc/1912.08175}}}.
  
  \bibitem{Kwidzinski:2020xyd} 
  N.~Kwidzinski, D.~Malafarina, J.~Ostrowski, W.~Piechocki and T.~Schmitz,
  ``{Hamiltonian formulation of dust cloud collapse,}''
  \href{http://arXiv.org/abs/gr-qc/2001.01903}{{\texttt{arXiv:gr-qc/2001.01903}}}.

\bibitem{Rovelli:2014cta}
C.~Rovelli and F.~Vidotto, ``{Planck stars},'' Int. J. Mod. Phys. {\bf D23}
  (2014), no.~12, 1442026,
\href{http://arXiv.org/abs/1401.6562}{{\texttt{arXiv:1401.6562}}}.

\bibitem{DeLorenzo:2014pta} 
  T.~De Lorenzo, C.~Pacilio, C.~Rovelli and S.~Speziale,
  ``{On the Effective Metric of a Planck Star,}''
  Gen.\ Rel.\ Grav.\  {\bf 47}, no. 4, 41 (2015)
  \href{http://arXiv.org/abs/1412.6015}{{\texttt{arXiv:1412.6015}}}.

\bibitem{Barrau:2014hda}
A.~Barrau and C.~Rovelli, ``{Planck star phenomenology},'' Phys. Lett. {\bf
  B739} (2014) 405--409,
\href{http://arXiv.org/abs/1404.5821}{{\texttt{arXiv:1404.5821}}}.

\bibitem{Barrau:2015uca}
A.~Barrau, B.~Bolliet, F.~Vidotto, and C.~Weimer, ``{Phenomenology of bouncing
  black holes in quantum gravity: a closer look},'' JCAP {\bf 1602} (2016),
  no.~02, 022,
\href{http://arXiv.org/abs/1507.05424}{{\texttt{arXiv:1507.05424}}}.

\bibitem{Barrau:2016fcg}
A.~Barrau, B.~Bolliet, M.~Schutten, and F.~Vidotto, ``{Bouncing black holes in
  quantum gravity and the Fermi gamma-ray excess},'' Phys. Lett. {\bf B772}
  (2017) 58--62,
\href{http://arXiv.org/abs/1606.08031}{{\texttt{arXiv:1606.08031}}}.

\bibitem{Rovelli:2017zoa}
C.~Rovelli, ``{Planck stars as observational probes of quantum gravity},'' Nat.
  Astron. {\bf 1} (2017) 0065,
\href{http://arXiv.org/abs/1708.01789}{{\texttt{arXiv:1708.01789}}}.

\bibitem{Barrau:2018kyv}
A.~Barrau, F.~Moulin, and K.~Martineau, ``{Fast radio bursts and the stochastic
  lifetime of black holes in quantum gravity},'' Phys. Rev. {\bf D97} (2018),
  no.~6, 066019,
\href{http://arXiv.org/abs/1801.03841}{{\texttt{arXiv:1801.03841}}}.

\bibitem{Rama:2019hdg} 
  S.~K.~Rama,
  ``Black hole or Fuzz ball or a Loop quantum star ? Assessing the fate of a massive collapsing star,''
  \href{http://arXiv.org/abs/1912.05300}{{\texttt{arXiv:1912.05300}}}.

\bibitem{Haggard:2014rza}
H.~M. Haggard and C.~Rovelli, ``{Quantum-gravity effects outside the horizon
  spark black to white hole tunneling},'' Phys. Rev. {\bf D92} (2015), no.~10,
  104020,
\href{http://arXiv.org/abs/1407.0989}{{\texttt{arXiv:1407.0989}}}.

\bibitem{Barcelo:2014npa}
C.~Barceló, R.~Carballo-Rubio, and L.~J. Garay, ``{Mutiny at the white-hole
  district},'' Int. J. Mod. Phys. {\bf D23} (2014), no.~12, 1442022,
\href{http://arXiv.org/abs/1407.1391}{{\texttt{arXiv:1407.1391}}}.

\bibitem{Barcelo:2014cla}
C.~Barcelo, R.~Carballo-Rubio, L.~J. Garay, and G.~Jannes, ``{The lifetime
  problem of evaporating black holes: mutiny or resignation},'' Class. Quant.
  Grav. {\bf 32} (2015), no.~3, 035012,
\href{http://arXiv.org/abs/1409.1501}{{\texttt{arXiv:1409.1501}}}.


\bibitem{DeLorenzo:2015gtx}
T.~De~Lorenzo and A.~Perez, ``{Improved Black Hole Fireworks: Asymmetric
  Black-Hole-to-White-Hole Tunneling Scenario},'' Phys. Rev. {\bf D93} (2016),
  no.~12, 124018,
\href{http://arXiv.org/abs/1512.04566}{{\texttt{arXiv:1512.04566}}}.

\bibitem{DAmbrosio:2018wgv} 
  F.~D'Ambrosio and C.~Rovelli,
  ``{How information crosses Schwarzschild’s central singularity,}''
  Class.\ Quant.\ Grav.\  {\bf 35}, no. 21, 215010 (2018)
  \href{http://arXiv.org/abs/1803.05015}{{\texttt{arXiv:1803.05015}}}.

 


\bibitem{Rovelli:2018cbg}
C.~Rovelli and P.~Martin-Dussaud, ``{Interior metric and ray-tracing map in the
  firework black-to-white hole transition},'' Class. Quant. Grav. {\bf 35}
  (2018), no.~14, 147002,
\href{http://arXiv.org/abs/1803.06330}{{\texttt{arXiv:1803.06330}}}.

\bibitem{Barcelo:2015uff}
C.~Barceló, R.~Carballo-Rubio, and L.~J. Garay, ``{Black holes turn white
  fast, otherwise stay black: no half measures},'' JHEP {\bf 01} (2016) 157,
\href{http://arXiv.org/abs/1511.00633}{{\texttt{arXiv:1511.00633}}}.

\bibitem{Barcelo:2015noa} 
  C.~Barceló, R.~Carballo-Rubio and L.~J.~Garay,
  ``Where does the physics of extreme gravitational collapse reside?,''
  Universe {\bf 2}, no. 2, 7 (2016)
  \href{http://arXiv.org/abs/1510.04957}{{\texttt{arXiv:1510.04957}}}.

\bibitem{Barcelo:2016hgb}
C.~Barceló, R.~Carballo-Rubio, and L.~J. Garay, ``{Exponential fading to white
  of black holes in quantum gravity},'' Class. Quant. Grav. {\bf 34} (2017),
  no.~10, 105007,
\href{http://arXiv.org/abs/1607.03480}{{\texttt{arXiv:1607.03480}}}.

\bibitem{Barcelo:2017lnx}
C.~Barceló, R.~Carballo-Rubio, and L.~J. Garay, ``{Gravitational wave echoes
  from macroscopic quantum gravity effects},'' JHEP {\bf 05} (2017) 054,
\href{http://arXiv.org/abs/1701.09156}{{\texttt{arXiv:1701.09156}}}.


\bibitem{Carballo-Rubio:2018pmi}
R.~Carballo-Rubio, F.~Di~Filippo, S.~Liberati, C.~Pacilio, and M.~Visser, ``{On
  the viability of regular black holes},'' JHEP {\bf 07} (2018) 023,
\href{http://arXiv.org/abs/1805.02675}{{\texttt{arXiv:1805.02675}}}.

 \bibitem{Bianchi:2018mml}
E.~Bianchi, M.~Christodoulou, F.~D'Ambrosio, H.~M. Haggard, and C.~Rovelli,
  ``{White Holes as Remnants: A Surprising Scenario for the End of a Black
  Hole},'' Class. Quant. Grav. {\bf 35} (2018), no.~22, 225003,
\href{http://arXiv.org/abs/1802.04264}{{\texttt{arXiv:1802.04264}}}.

   \bibitem{Rovelli:2018okm} 
  C.~Rovelli and F.~Vidotto,
  ``{Small black/white hole stability and dark matter,}''
  Universe {\bf 4}, no. 11, 127 (2018)
     \href{http://arXiv.org/abs/1805.03872}{{\texttt{arXiv:1805.03872}}}
     
     \bibitem{Rovelli:2018hbk}   
  C.~Rovelli and F.~Vidotto,
  ``{White-hole dark matter and the origin of past low-entropy,}''
     \href{http://arXiv.org/abs/1804.04147}{{\texttt{arXiv:1804.04147}}}

    \bibitem{Martin-Dussaud:2019wqc}   
  P.~Martin-Dussaud and C.~Rovelli,
  ``{Evaporating black-to-white hole,}''
  Class.\ Quant.\ Grav.\  {\bf 36}, no. 24, 245002 (2019)
    \href{http://arXiv.org/abs/1905.07251}{{\texttt{arXiv:1905.07251}}}.
   
    
      \bibitem{Baccetti:2018qrp} 
  V.~Baccetti, S.~Murk and D.~R.~Terno,
  ``{Black hole evaporation and semiclassical thin shell collapse,}''
  Phys.\ Rev.\ D {\bf 100}, no. 6, 064054 (2019)
        \href{http://arXiv.org/abs/1812.07727}{{\texttt{arXiv:1812.07727}}}.
  
  \bibitem{Baccetti:2016lsb} 
  V.~Baccetti, R.~B.~Mann and D.~R.~Terno,
  ``{Role of evaporation in gravitational collapse,}''
  Class.\ Quant.\ Grav.\  {\bf 35}, no. 18, 185005 (2018)
     \href{http://arXiv.org/abs/1610.07839}{{\texttt{arXiv:1610.07839}}}
  
  \bibitem{Baccetti:2019mab} 
  V.~Baccetti, R.~B.~Mann and D.~R.~Terno,
  ``{Trapped surfaces, energy conditions, and horizon avoidance in spherically-symmetric collapse,}''
       \href{http://arXiv.org/abs/1904.00506}{{\texttt{arXiv:1904.00506}}}
       
       \bibitem{Ho:2019pjr} 
  P.~M.~Ho, Y.~Matsuo and Y.~Yokokura,
  ``{An Analytic Description of Semi-Classical Black-Hole Geometry,}''
         \href{http://arXiv.org/abs/1912.12855}{{\texttt{arXiv:1912.12855}}}
  
  \bibitem{Ho:2019qiu} 
  P.~M.~Ho, Y.~Matsuo and Y.~Yokokura,
  ``{Distance between collapsing matter and trapping horizon in evaporating black holes,}''
    \href{http://arXiv.org/abs/1912.12863}{{\texttt{arXiv:1912.12863}}}
    
           \bibitem{Goswami:2005fu} 
  R.~Goswami, P.~S.~Joshi and P.~Singh,
  ``{Quantum evaporation of a naked singularity,}''
  Phys.\ Rev.\ Lett.\  {\bf 96}, 031302 (2006)
    \href{http://arXiv.org/abs/0506129}{{\texttt{arXiv:0506129}}}.
    
     \bibitem{Corichi:2015xia} 
  A.~Corichi and P.~Singh,
  ``{Loop quantization of the Schwarzschild interior revisited,}''
  Class.\ Quant.\ Grav.\  {\bf 33}, no. 5, 055006 (2016)
  \href{http://arXiv.org/abs/1506.08015}{{\texttt{arXiv:1506.08015}}}.

\bibitem{Olmedo:2017lvt} 
  J.~Olmedo, S.~Saini and P.~Singh,
  ``{From black holes to white holes: a quantum gravitational, symmetric bounce,}''
  Class.\ Quant.\ Grav.\  {\bf 34}, no. 22, 225011 (2017)
  \href{http://arXiv.org/abs/1707.07333}{{\texttt{arXiv:1707.07333}}}.

\bibitem{Ashtekar:2018cay} 
  A.~Ashtekar, J.~Olmedo and P.~Singh,
  ``{Quantum extension of the Kruskal spacetime,}''
  Phys.\ Rev.\ D {\bf 98}, no. 12, 126003 (2018)
  \href{http://arXiv.org/abs/1806.02406}{{\texttt{arXiv:1806.02406}}}.
  
  \bibitem{Bodendorfer:2019cyv} 
  N.~Bodendorfer, F.~M.~Mele and J.~Munch,
  ``{Effective Quantum Extended Spacetime of Polymer Schwarzschild Black Hole,}''
  Class.\ Quant.\ Grav.\  {\bf 36}, no. 19, 195015 (2019)
  \href{http://arXiv.org/abs/1902.04542}{{\texttt{arXiv:1902.04542}}}.
  
  \bibitem{Bodendorfer:2019nvy} 
  N.~Bodendorfer, F.~M.~Mele and J.~Munch,
  ``{(b,v)-type variables for black to white hole transitions in effective loop quantum gravity,}''
    \href{http://arXiv.org/abs/1911.12646}{{\texttt{arXiv:1911.12646}}}.

  
  \bibitem{Bodendorfer:2019jay} 
  N.~Bodendorfer, F.~M.~Mele and J.~Munch,
  ``{Mass and Horizon Dirac Observables in Effective Models of Quantum Black-to-White Hole Transition,}''
    \href{http://arXiv.org/abs/1912.00774}{{\texttt{arXiv:1912.00774}}}.

\bibitem{Assanioussi:2019twp} 
  M.~Assanioussi, A.~Dapor and K.~Liegener,
  ``Perspectives on the dynamics in a loop quantum gravity effective description of black hole interiors,''
   \href{http://arXiv.org/abs/1908.05756}{{\texttt{arXiv:1908.05756}}}. 
    
        \bibitem{Alesci:2018loi} 
  E.~Alesci, S.~Bahrami and D.~Pranzetti,
  ``{Quantum evolution of black hole initial data sets: Foundations,}''
  Phys.\ Rev.\ D {\bf 98}, no. 4, 046014 (2018)
      \href{http://arXiv.org/abs/1807.07602}{{\texttt{arXiv:1807.07602}}}.
  
  \bibitem{Alesci:2019pbs}   
  E.~Alesci, S.~Bahrami and D.~Pranzetti,
  ``{Quantum gravity predictions for black hole interior geometry,}''
  Phys.\ Lett.\ B {\bf 797}, 134908 (2019)
      \href{http://arXiv.org/abs/1807.07602}{{\texttt{arXiv:1807.07602}}}.
    
  \bibitem{BenAchour:2018khr} 
  J.~Ben Achour, F.~Lamy, H.~Liu and K.~Noui,
  ``{Polymer Schwarzschild black hole: An effective metric,}''
  EPL {\bf 123}, no. 2, 20006 (2018)
   \href{http://arXiv.org/abs/1803.01152}{{\texttt{arXiv:1803.01152}}}.
  
  \bibitem{Bojowald:2018xxu} 
  M.~Bojowald, S.~Brahma and D.~h.~Yeom,
  ``{Effective line elements and black-hole models in canonical loop quantum gravity,}''
  Phys.\ Rev.\ D {\bf 98}, no. 4, 046015 (2018)
   \href{http://arXiv.org/abs/1803.01119}{{\texttt{arXiv:1803.01119}}}.
   
   \bibitem{Aruga:2019dwq} 
  D.~Aruga, J.~Ben Achour and K.~Noui,
  ``{Deformed General Relativity and Quantum Black Holes Interior,}''
   \href{http://arXiv.org/abs/1912.02459}{{\texttt{arXiv:1912.02459}}}.
  
   
   \bibitem{Bojowald:2015zha} 
  M.~Bojowald, S.~Brahma and J.~D.~Reyes,
  ``{Covariance in models of loop quantum gravity: Spherical symmetry,}''
  Phys.\ Rev.\ D {\bf 92}, no. 4, 045043 (2015)
     \href{http://arXiv.org/abs/1507.00329}{{\texttt{arXiv:1507.00329}}}.

\bibitem{Brahma:2014gca} 
  S.~Brahma,
  ``Spherically symmetric canonical quantum gravity,''
  Phys.\ Rev.\ D {\bf 91}, no. 12, 124003 (2015)
  \href{http://arXiv.org/abs/1411.3661}{{\texttt{arXiv:1411.3661}}}.
  
  \bibitem{BenAchour:2016brs} 
  J.~Ben Achour, S.~Brahma and A.~Marciano,
  ``{Spherically symmetric sector of self dual Ashtekar gravity coupled to matter: Anomaly-free algebra of constraints with holonomy corrections,}''
  Phys.\ Rev.\ D {\bf 96}, no. 2, 026002 (2017)
    \href{http://arXiv.org/abs/1608.07314}{{\texttt{arXiv:1608.07314}}}.
    
    \bibitem{BenAchour:2017jof} 
  J.~Ben Achour and S.~Brahma,
  ``{Covariance in self dual inhomogeneous models of effective quantum geometry: Spherical symmetry and Gowdy systems,}''
  Phys.\ Rev.\ D {\bf 97}, no. 12, 126003 (2018)
      \href{http://arXiv.org/abs/1712.03677}{{\texttt{arXiv:1712.03677}}}.
      
      \bibitem{Bojowald:2019fkv} 
  M.~Bojowald, S.~Brahma, D.~Ding and M.~Ronco,
  ``{Deformed covariance in spherically symmetric vacuum models of loop quantum gravity: Consistency in Euclidean and self-dual gravity,}''
  Phys.\ Rev.\ D {\bf 101}, no. 2, 026001 (2020)
        \href{http://arXiv.org/abs/1910.10091}{{\texttt{arXiv:1910.10091}}}.

\bibitem{Christodoulou:2016vny} 
  M.~Christodoulou, C.~Rovelli, S.~Speziale and I.~Vilensky,
  ``{Planck star tunneling time: An astrophysically relevant observable from background-free quantum gravity,}''
  Phys.\ Rev.\ D {\bf 94}, no. 8, 084035 (2016)
  \href{http://arXiv.org/abs/1605.05268}{{\texttt{arXiv:1605.05268}}}.
  

  
  \bibitem{Christodoulou:2018ryl} 
  M.~Christodoulou and F.~D'Ambrosio,
  ``{Characteristic Time Scales for the Geometry Transition of a Black Hole to a White Hole from Spinfoams,}''
    \href{http://arXiv.org/abs/1801.03027}{{\texttt{arXiv:1801.03027}}}.
     
  
  \bibitem{Marolf:2017jkr} 
  D.~Marolf,
  ``{The Black Hole information problem: past, present, and future,}''
  Rept.\ Prog.\ Phys.\  {\bf 80}, no. 9, 092001 (2017)
    \href{http://arXiv.org/abs/1703.02143}{{\texttt{arXiv:1703.02143}}}.
    
    \bibitem{Compere:2019ssx} 
  G.~Compère,
  ``{Are quantum corrections on horizon scale physically motivated?,}''
  Int.\ J.\ Mod.\ Phys.\ D {\bf 28}, no. 14, 1930019 (2019)
   \href{http://arXiv.org/abs/1902.04504}{{\texttt{arXiv:1902.04504}}}.
   
   \bibitem{Rovelli:2019cik} 
  C.~Rovelli,
  ``{The Subtle Unphysical Hypothesis of the Firewall Theorem,}''
  Entropy {\bf 21}, no. 9, 839 (2019)
  \href{http://arXiv.org/abs/1902.03631}{{\texttt{arXiv:1902.03631}}}.


\bibitem{Brandenberger:2016vhg}
R.~Brandenberger and P.~Peter, ``{Bouncing Cosmologies: Progress and
  Problems},'' Found. Phys. {\bf 47} (2017), no.~6, 797--850,
\href{http://arXiv.org/abs/1603.05834}{{\texttt{arXiv:1603.05834}}}.

\bibitem{Battefeld:2014uga}
D.~Battefeld and P.~Peter, ``{A Critical Review of Classical Bouncing
  Cosmologies},'' Phys. Rept. {\bf 571} (2015) 1--66,
\href{http://arXiv.org/abs/1406.2790}{{\texttt{arXiv:1406.2790}}}.

\bibitem{Ashtekar:2011ni}
A.~Ashtekar and P.~Singh, ``{Loop Quantum Cosmology: A Status Report},'' Class.
  Quant. Grav. {\bf 28} (2011) 213001,
\href{http://arXiv.org/abs/1108.0893}{{\texttt{arXiv:1108.0893}}}.

\bibitem{Bojowald:2005qw}
M.~Bojowald, R.~Goswami, R.~Maartens, and P.~Singh, ``{A Black hole mass
  threshold from non-singular quantum gravitational collapse},'' Phys. Rev.
  Lett. {\bf 95} (2005) 091302,
\href{http://arXiv.org/abs/gr-qc/0503041}{{\texttt{arXiv:gr-qc/0503041}}}.

\bibitem{Tavakoli:2013lga}
Y.~Tavakoli, J.~Marto, and A.~Dapor, ``{Dynamics of apparent horizons in
  quantum gravitational collapse},'' Springer Proc. Math. Stat. {\bf 60} (2014)
  427--431,
\href{http://arXiv.org/abs/1306.3458}{{\texttt{arXiv:1306.3458}}}.

\bibitem{Tavakoli:2013rna}
Y.~Tavakoli, J.~Marto, and A.~Dapor, ``{Semiclassical dynamics of horizons in
  spherically symmetric collapse},'' Int. J. Mod. Phys. {\bf D23} (2014),
  no.~7, 1450061,
\href{http://arXiv.org/abs/1303.6157}{{\texttt{arXiv:1303.6157}}}.

\bibitem{Bambi:2013caa}
C.~Bambi, D.~Malafarina, and L.~Modesto, ``{Non-singular quantum-inspired
  gravitational collapse},'' Phys. Rev. {\bf D88} (2013) 044009,
\href{http://arXiv.org/abs/1305.4790}{{\texttt{arXiv:1305.4790}}}.

\bibitem{Liu:2014kra}
Y.~Liu, D.~Malafarina, L.~Modesto, and C.~Bambi, ``{Singularity avoidance in
  quantum-inspired inhomogeneous dust collapse},'' Phys. Rev. {\bf D90} (2014),
  no.~4, 044040,
\href{http://arXiv.org/abs/1405.7249}{{\texttt{arXiv:1405.7249}}}.

\bibitem{These}
S.~Campbell, ``{Models of Non-Singular Gravitational collapse},'' Ph.D Thesis {\bf
  London SW7 2AZ}
(2014).

\bibitem{Israel:1966rt}
W.~Israel, ``{Singular hypersurfaces and thin shells in general relativity},''
  Nuovo Cim. {\bf B44S10} (1966) 1.
[Nuovo Cim.B44,1(1966)].

\bibitem{Barrabes:1991ng}
C.~Barrabes and W.~Israel, ``{Thin shells in general relativity and cosmology:
  The Lightlike limit},'' Phys. Rev. {\bf D43} (1991)
1129--1142.

\bibitem{Easson:2011zy} 
  D.~A.~Easson, I.~Sawicki and A.~Vikman,
  ``{G-Bounce,}''
  JCAP {\bf 1111}, 021 (2011)
  \href{http://arXiv.org/abs/1109.1047}{{\texttt{arXiv:1109.1047}}}
  

\bibitem{Qiu:2011cy} 
  T.~Qiu, J.~Evslin, Y.~F.~Cai, M.~Li and X.~Zhang,
  ``{Bouncing Galileon Cosmologies,}''
  JCAP {\bf 1110}, 036 (2011)
     \href{http://arXiv.org/abs/1108.0593}{{\texttt{arXiv:1108.0593}}}.
     
       \bibitem{Cai:2012va} 
  Y.~F.~Cai, D.~A.~Easson and R.~Brandenberger,
  ``{Towards a Nonsingular Bouncing Cosmology,}''
  JCAP {\bf 1208}, 020 (2012)
   \href{http://arXiv.org/abs/1206.2382}{{\texttt{arXiv:1206.2382}}}.

 \bibitem{Li:2014era} 
  C.~Li, R.~H.~Brandenberger and Y.~K.~E.~Cheung,
  ``{Big-Bounce Genesis,}''
  Phys.\ Rev.\ D {\bf 90}, no. 12, 123535 (2014)
    \href{http://arXiv.org/abs/1403.5625}{{\texttt{arXiv:1403.5625}}}.
  
  \bibitem{Qiu:2015nha} 
  T.~Qiu and Y.~T.~Wang,
  ``{G-Bounce Inflation: Towards Nonsingular Inflation Cosmology with Galileon Field,}''
  JHEP {\bf 1504}, 130 (2015)
    \href{http://arXiv.org/abs/1501.03568}{{\texttt{arXiv:1501.03568}}}.
  
  
  \bibitem{Kolevatov:2017voe} 
  R.~Kolevatov, S.~Mironov, N.~Sukhov and V.~Volkova,
  ``{Cosmological bounce and Genesis beyond Horndeski,}''
  JCAP {\bf 1708}, 038 (2017)
   \href{http://arXiv.org/abs/1705.06626}{{\texttt{arXiv:1705.06626}}}.
   
   \bibitem{Ijjas:2016tpn} 
  A.~Ijjas and P.~J.~Steinhardt,
  ``{Classically stable nonsingular cosmological bounces,}''
  Phys.\ Rev.\ Lett.\  {\bf 117}, no. 12, 121304 (2016)
     \href{http://arXiv.org/abs/1705.06626}{{\texttt{arXiv:1606.08880 }}}.
     
       \bibitem{Dobre:2017pnt} 
  D.~A.~Dobre, A.~V.~Frolov, J.~T.~G.~Ghersi, S.~Ramazanov and A.~Vikman,
  ``{Unbraiding the Bounce: Superluminality around the Corner,}''
  JCAP {\bf 1803}, 020 (2018)
       \href{http://arXiv.org/abs/1712.10272}{{\texttt{arXiv:1712.10272}}}.
     
       \bibitem{Gao:2017ihf} 
  C.~Gao, Y.~Lu, Y.~G.~Shen and V.~Faraoni,
  ``{Pulsation of black holes,}''
  Gen.\ Rel.\ Grav.\  {\bf 50}, no. 1, 15 (2018)
      \href{http://arXiv.org/abs/1706.08009}{{\texttt{arXiv:1706.08009}}}.
   
      \bibitem{paperUS}  
  J.~Ben Achour and J.~P.~Uzan,
  ``{Bouncing compact objects II: Effective theory of a pulsating Planck star,}''
        \href{http://arXiv.org/abs/2001.06153}{{\texttt{arXiv:2001.06153}}}.
  
  
  \bibitem{Ashtekar:2006es}  
  A.~Ashtekar, T.~Pawlowski, P.~Singh and K.~Vandersloot,
  ``{Loop quantum cosmology of k=1 FRW models,}''
  Phys.\ Rev.\ D {\bf 75}, 024035 (2007)
    \href{http://arXiv.org/abs/0612104}{{\texttt{arXiv:0612104}}}.
    
    \bibitem{Szulc:2006ep} 
  L.~Szulc, W.~Kaminski and J.~Lewandowski,
  ``{Closed FRW model in Loop Quantum Cosmology,}''
  Class.\ Quant.\ Grav.\  {\bf 24}, 2621 (2007)
    \href{http://arXiv.org/abs/0612101}{{\texttt{arXiv:0612101}}}.

\bibitem{Corichi:2011pg} 
  A.~Corichi and A.~Karami,
  ``{Loop quantum cosmology of k=1 FRW: A tale of two bounces,}''
  Phys.\ Rev.\ D {\bf 84}, 044003 (2011)
  \href{http://arXiv.org/abs/1105.3724}{{\texttt{arXiv:1105.3724}}}.
  
  \bibitem{Dupuy:2016upu} 
  J.~L.~Dupuy and P.~Singh,
  ``{Implications of quantum ambiguities in $k$=1 loop quantum cosmology: distinct quantum turnarounds and the super-Planckian regime,}''
  Phys.\ Rev.\ D {\bf 95}, no. 2, 023510 (2017)
    \href{http://arXiv.org/abs/1608.07772}{{\texttt{arXiv:1608.07772}}}.
   



  \bibitem{Carballo-Rubio:2019nel} 
  R.~Carballo-Rubio, F.~Di Filippo, S.~Liberati and M.~Visser,
  ``{Opening the Pandora's box at the core of black holes,}''
    \href{http://arXiv.org/abs/1908.03261}{{\texttt{arXiv:1908.03261}}}.
    
    \bibitem{Carballo-Rubio:2019fnb} 
  R.~Carballo-Rubio, F.~Di Filippo, S.~Liberati and M.~Visser,
  ``{Geodesically complete black holes,}''
   \href{http://arXiv.org/abs/1911.11200}{{\texttt{arXiv:1911.11200}}}.
  
     \bibitem{futureUS} 
  J.~Ben Achour, S.~Brahma, S. Mukohyama and J-P.~Uzan,  ``{Black-to-White hole from matter collapse: A new model-independent constraint,}''
  "to appear" (2020)
   
   \bibitem{Brahma:2018cgr} 
  S.~Brahma and D.~h.~Yeom,
  ``Effective black-to-white hole bounces: The cost of surgery,''
  Class.\ Quant.\ Grav.\  {\bf 35}, no. 20, 205007 (2018)
  \href{http://arXiv.org/abs/1804.02821}{{\texttt{arXiv:1804.02821}}}.
   

  \bibitem{Rovelli:2013zaa} 
  C.~Rovelli and E.~Wilson-Ewing,
  ``Why are the effective equations of loop quantum cosmology so accurate?,''
  Phys.\ Rev.\ D {\bf 90}, no. 2, 023538 (2014)
   \href{http://arXiv.org/abs/1310.8654}{{\texttt{arXiv:1310.8654}}}.
  
  \bibitem{Bojowald:2015fla} 
  M.~Bojowald and S.~Brahma,
  ``Minisuperspace models as infrared contributions,''
  Phys.\ Rev.\ D {\bf 93}, no. 12, 125001 (2016)
  \href{http://arXiv.org/abs/1509.00640}{{\texttt{arXiv:1509.00640}}}.
  

  


  


  
\end{thebibliography}
\end{document}